# Crown-structured optical harmonics

David L. Andrews,[1]* Henryk Haniewicz,[1] and Enrique J. Galvez[2]

[1] *University of East Anglia, Norwich Research Park, Norwich NR4 7TJ, UK*
[2] *Colgate University, Hamilton, New York 13346, USA*

At levels of laser intensity below threshold for multiphoton ionization, the parametric generation of optical harmonics in gases and other isotropic media is subject to selection rules with origins in angular momentum conservation. The recently developed optics of vector polarization modes provides an unprecedented opportunity to exploit these principles in the production of high harmonic beams with distinctive forms of transverse intensity profile, comprising discrete sub-wavelength filaments in crown-like arrays. A detailed analysis of the fundamental electrodynamics elicits the mechanism, and delivers results illustrating the transverse structures and spatial dimensions of harmonic output that can be achieved.

High harmonic generation is one of the most important optical techniques for producing attosecond pulses of UV and soft x-ray light, offering capabilities for extremely high spatial and temporal resolution.[1] As is well known, the coherent generation of high or low order optical harmonics in regular gases, in other isotropic or centrosymmetric solids or liquids,[2, 3] is subject to powerful selection rules that preclude the coherent generation of even order harmonic frequencies. While issues of fundamental symmetry lie behind this condition,[4] phase-matching is also important; in contrast, incoherent harmonic scattering allows only for the production of very weak signals. Considerations of photon angular momentum also dictate that generation of coherent harmonics from a circularly polarized input is entirely forbidden in the gas phase, or indeed in other isotropic media.[5, 6] To circumvent this rule for high harmonic production of circular polarizations typically in the extreme ultraviolet, it is usually necessary to deploy pump beams with a significant degree of non-collinearity.[7] Alternatively, bichromatic fields with counter-rotating polarizations can be used, to enforce compliance with the selection rules and so produce circularly polarized high harmonic emission.[8]

It has recently been shown possible to generate extremely tight spots of light by producing petal-like superpositions of co-propagating beams having opposite orbital angular momentum.[9] Progressing the continued development of needle-like beams,[10] it has also been shown that an individual optical needle of wavelength magnitude can be produced by tight focusing of a partially coherent input.[11] We can now show that the recently developed optics of vector polarization modes[12] provides another, very different opportunity to creatively exploit the symmetry principles underlying coherent harmonic generation, in the production of strong odd-harmonic output beams with highly distinctive, typically crown-shaped forms of intensity profile. As will be shown, the transverse spatial profile of these coherent harmonics, created from single-beam input, comprise rings of needle-like filaments. Individually, these high-intensity filaments are characterized by an unusually narrow spatial profile, facilitating an increasingly enhanced spatial resolution as the order of the harmonic increases. Such features may prove especially important for efficient coupling into nano-structures and waveguides. The results for high order harmonic generation also offer particular significance for spatially sensitive attosecond instrumentation.[13] It is the purpose of this Letter to explain the mechanism that underpins the crown formation, and to exhibit the kinds of spatially structured harmonic output that now become possible.

Circularly polarized photons convey spin angular momentum $S$ as single units of the reduced Planck's constant $\hbar$. In the optics convention, the sign is positive for left-handed polarization, negative for right. All other states of polarization relate to superpositions of left and right polarizations: a resolution of plane polarized light into circular components collapses its state into one or other handedness with equal probability. The preclusion of harmonics of even order $m$, for coherent forward emission in an isotropic medium, reflects the fact that no even combination of ±1 can deliver a single unit of angular momentum to a single output photon, to satisfy angular momentum conservation. For odd $m$ harmonics, it is always possible to satisfy this conservation law – provided the input is not a pure circular polarization, i.e. it cannot consist of solely photons with the same circular handedness. In cases where the input does have a pure circular polarization it is impossible to reconcile the additive spin input with the output of just one unit. The only direct way around this rule is to double-up on all the interactions – for example using a six-wave mixing method to produce frequency-doubled output (with two output photons emerging for every four photons of input).[14-17] All of these principles can be elicited from a mathematical analysis of the isotropic tensors featured in the electrodynamics of optical harmonic generation.[6]

The local efficiency of conversion to any optical harmonic of order $m$ depends in general on the instantaneous intensity, $\bar{I}$, the state of polarization and photon statistics of the pump radiation – in the latter case specifically $g^{(m)}$, the degree of $m^{\text{th}}$ order coherence.[18] In a vector polarization structured beam, each of these parameters may vary across the beam profile. However, since such beams are generally produced by the passage of a conventional beam through passive optical elements, the coherence factors at least may be assumed essentially constant. For the most common forms of vector polarization structure, the intensity of light has a radial distribution and, in the simplest cases, the local state of polarization is furthermore dependent only on the azimuthal angle around the beam axis. More complicated structures are possible; the present analysis serves to introduce the operation of the mechanism for crown harmonic formation.

The detailed form of expression for the harmonic intensity is dictated by the fundamental electrodynamics of the harmonic conversion process; phase-matching is responsible for sustaining optical coherence in the output. For conversion in an isotropic medium, coherence signifies that, in the theory, a rotational averaging must be mathematically performed upon the quantum amplitude (as opposed to its implementation on the rate equation).[19] The full theory gives the following dependence of the harmonic intensity $I_{m\text{HG}}$ on the input:

$$I_{m\text{HG}} \sim \left(\bar{I}/I_0\right)^m \left|\chi^{(m)}\right|^2 g^{(m)} \left|(\mathbf{e}.\mathbf{e})^{(m-1)/2}(\mathbf{e}.\bar{\mathbf{e}}')\right|^2. \quad (1)$$

Here the unit polarization vectors are denoted by $\mathbf{e}$ for the input, $\mathbf{e}'$ for the output; each may be freely represented by any point on the Poincaré or Bloch sphere. For polarization-structured light, $\mathbf{e}$, $\mathbf{e}'$, $\bar{I}$ – and hence also $I_{m\text{HG}}$, are in general locally dependent on position with respect to the beam axis; as such, they are conveniently expressed in terms of transverse polar coordinates $(r,\phi)$. Furthermore, the scalar quantity $\chi^{(m)}$ is the isotropic part of the nonlinear optical susceptibility tensor for $m$-harmonic generation, and $I_0$ is a non-local measure of irradiance at which perturbation theory ceases to operate, due to the onset of high-field effects such as ionization. It may be noted that the exact power dependence on irradiance, exhibited in equation (1), need not be assumed to signify direct (as opposed to cascaded) multiphoton annihilation of the input wave: it reflects a condition for $m$ photons of the input to be simultaneously present within a region containing an atomic or materially distinct, representative unit of the nonlinear optical medium.[20] Despite this power dependence, the involvement of the nonlinear susceptibility commonly tempers, to a significant degree, any anticipated decline of harmonic intensity with increasing order $m$; as is well known, a plateau region with very little variation of intensity between successive odd harmonics can ultimately be produced under ionizing conditions.[21, 22]

Straightforward analysis shows that, summing the result from equation (1) over output polarizations, the following expression is obtained for the transverse distribution of each harmonic;

$$I_{m\text{HG}}(r,\phi) \sim \left(\bar{I}(r,\phi)/I_0\right)^m \sin^{m-1}\{2\chi(r,\phi)\}, \quad (2)$$

where $\chi$ (no superscript) here has its usual meaning as a coordinate on the Poincaré sphere, with $\chi(r,\phi)$ representing a generally intricate form of variation across the beam cross-section. Furthermore, in the most common forms of implementation the intensity distribution is cylindrical, and the polarization structures are radial,[23] so that $\bar{I}^m$ then depends explicitly only on the radial coordinate $r$, and the spatial distribution of the harmonics reduces to $\bar{I}^m(r)\left(1-\varepsilon^2(\phi)\right)^{m-1}$, where the ellipticity $\varepsilon$ is a function of the azimuthal angle $\phi$.

The production of vector polarization beams is achievable by various means. One well-proven and studied methods involves the superposition of Laguerre-Gaussian vortex modes[24] – those in turn being produced from the passage of a conventional laser output through optical elements such as spatial light modulators,[25] cascaded $J$-plates,[26, 27] or a more extensive combination of such elements that can produce almost 90% conversion efficiencies.[28] It proves that a close match in magnitude between topological charges $\ell_1$ and $\ell_2$ of opposite sign, plus polarization modulation, most effectively generates polarization landscapes with substantial variations in polarization ellipticity, as illustrated in Fig. 1. In particular, when $|\ell_1 + \ell_2| = 0$ there is no residue of orbital angular momentum in the constructed beam, that might otherwise contribute to the harmonic frequency conversion.[29]

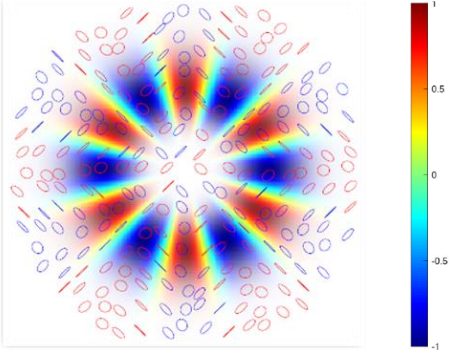

FIG. 1. Polarization landscape of a Poincaré beam obtained by combining Laguerre-Gaussian (LG) modes of topological charge $\ell = +3$ and -3 in non-separable superposition with linear polarization states. Colors code for the ellipticity, from red for right-handed to blue for left-handed circularity: the depth of color indicates intensity.

In the process of harmonic emission, using such a constructed beam as the pump, input photons from closely proximal regions of the beam are consumed in each conversion event. The validity of this assumption has recently been verified by explicit analysis and calculation for nonlinear optical processes of a generic parametric form.[30, 31] From this work it has been shown that processes involving more than one optical center, representing a discrete unit of the nonlinear optical susceptibility, will typically contribute no more than a few percent to the output. The requisite number of impinging photons for each individual output photon are mostly converted through interaction with in material whose dimensions are substantially smaller than the beam waist.

From the present analysis, it emerges that the output produced by harmonic generation from any such structured input can take a highly intricate form. For the case of the input exhibited in Fig. 1, with $S_{12}$ Schoenflies symmetry, the output produced by high-harmonic conversion consists of a ring of 12 intensity spikes of increasing sharpness as the order of the harmonic increases. In the sample results shown in Fig. 2, delivered from Eq. (2), the transverse intensity profiles are exhibited on an arbitrary scale, for three representative orders of harmonic; $m = 3, 19, 31$. In practice, these and all other odd-order harmonics will be superimposed with relative intensities simply determined by the magnitudes of the corresponding $(m+1)$-order susceptibilities of the optically nonlinear medium. Notably, the crown-like spatial profiles of the higher harmonics become extremely sharp – a feature that may be effectively exploited with a suitable cut-off filter.

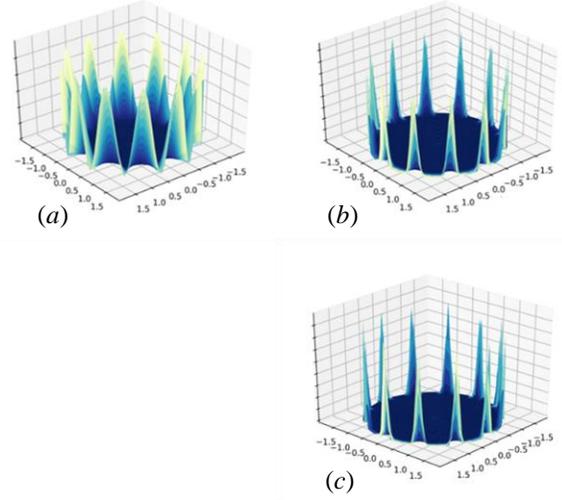

FIG/ 2. Transverse $(x, y)$ intensity profiles of the harmonics 3, 19 and 31 delivered by high-harmonic generation with the $\ell = 6$ Poincaré beam input whose transverse polarization landscape is shown in Fig. 1. The displacement distances on the $x$ and $y$ axes are in units of $w$, the radial distance of maximum intensity with a 'donut beam' input.

To gauge the relative width of each filament (intensity spike) in a typical case, we first consider one of the vector vortex structures featured in recent work by Huang et al.[27] For modes with a topological charge of unity, as for example is the case for those illustrated in Figs 7 (c) and (d) of that study, a rotary orbit within the beam produces the polarization effect of circumnavigating the Poincaré sphere in a polar orbit (akin to the effect of propagating through full-wave plate). There are, in the physical space orthogonal to the beam axis, two positions of pure circular polarization with opposite sign. Given an input of this form with a paraxial $LG_{1,0}$ field distribution and beam width parameter $w$, the $(r,\phi)$ transverse intensity profile resulting from conversion to an order $m$ optical harmonic takes the following form;

$$I_{mHG}(r,\phi) \sim \left[ r\exp\left(-(r/w)^2\right) \right]^{2m} (\cos\phi)^{(m-1)}. \qquad (3)$$

It is expedient to adopt a simple model exemplified by this case. For any such input with topological charge $\ell$, and polarizations that are independent of the radial coordinate $r$, the angular distribution of intensity becomes expressible as $(\cos\ell\phi)^{(m-1)/2}$ constituting a ring of $2\ell$ intensity filaments, equally spaced around the beam axis. From this expression, we can readily estimate a degree of effective azimuthal confinement for each filament. Each has an azimuthal FWHM (full width at half maximum) linewidth given by $(2/\ell)\cos^{-1}(2^{-2/(m-1)})$, and the corresponding circumferential width at the most intense radial position is therefore $(4\pi w/2^{1/2}\ell)\cos^{-1}(2^{-2/(m-1)})$. For example, the case illustrated in Fig. 2(c), with $\ell = 6$ and $m = 31$, produces a FWHM spatial width for each filament of $0.10^c$, representing 1.6% of the ring circumference given by $2^{1/2}\pi w$. The radial FWHM according to equation (3) is $0.16w$. This is to be contrasted with Fig. 2(a), the case of the lowest harmonic, $m = 3$, for which the result with $\ell = 6$ is $0.35^c$; this represents 5.6% of the ring per filament, and accordingly a not-significant overlap between successive filaments around the ring of radial width $0.48w$.

In practice, when a spectrally broad distribution of discrete high harmonic frequencies is produced, the spatial intensity profile of each filament (whose number depends only on the input polarization structure) will be an intensity-weighted superposition of contributions for a range of $m$ values. If, for example, we were to assume a flat intensity distribution for odd harmonics ranging from, say, $m = 19$ to $m = 39$ (filtering off the lower harmonics and accounting for an effective cut-off at the short-wavelength end) then for $\ell = 6$, each filament comprising these superimposed harmonics will have a FWHM of $0.11^c$, representing an annular dimension of $0.47w$. For $\ell = 12$ the corresponding result is halved. The net radial FWHM under these conditions is $0.17w$. In each case it is notable that the harmonic filaments are of substantially smaller dimensions than the output beam waist; the effect enhances the natural linewidth reduction associated with multiphoton excitation. Moreover, the symmetry of the distributed array is entirely determined by the polarization structure of the input.

Deploying these principles enables the structures of intricate polarization fields to be directly translated into intensity profiles comprising filaments of substantially sub-wavelength dimension. Important work on photonic metasurfaces has recently shown that for harmonic conversion in media with any finite degree $n$ of axial rotational symmetry, the operation of generalized rules for spin angular momentum coupling permits a reconciliation of circularly polarized input and output through uptake by the medium of $n\hbar$ in each conversion event.[32] Recognition of this principle readily enables the scale and efficiency of the described mechanism to be modified for media of various kinds of condensed phase medium of well-defined structural symmetry.

The underlying principle of our study is that for coherent parametric processes in isotropic or centrosymmetric media, there is no exchange of angular momentum between the radiation field and matter. This is in principle applicable to other kinds of optical parametric process. Limitations to the scope of the methods described here will occur at the highest levels of laser input, under strong-field conditions. Where harmonic conversion occurs in gas plasma, for example, both odd and even harmonics may appear: spontaneous emission may contribute to the output as new mechanisms come into play. It has long been known that the local production of static electric fields produces an additional interaction, engaging in the process of harmonic conversion and so permitting the production of even harmonics.[33] In a recent more extensive analysis it has notably been shown that quantum field methods lead to results that no longer comply with the predictions of semiclassical methods.[34]

The present analysis is directly applicable to systems that are essentially isotropic over the scale and volume within which harmonic conversion occurs. While it is well known that even harmonics are generally forbidden in such media, (and also those in which there is rotational symmetry of even order along the direction of propagation),[35] Saito et al. have recently provided a more detailed investigation of the symmetry conditions for high harmonic generation, given circularly polarized input, in crystalline solids of anisotropic form.[36] Their work shows that propagation along an axis of threefold symmetry does permit even harmonics to appear; in general, a lack of inversion symmetry undermines the validity of the rule over converting circular polarizations. This study affords a basis for experimentally investigating the transverse intensity profile of harmonics that may be produced on passage of vector polarization beams through uniaxial and biaxial optical materials.

In conclusion, it has been shown by the present study that there is now a rich scope to produce intricately structured beams of short-wavelength light, at high intensities. The results of the analysis illustrate the capacity of the vector polarization harmonic mechanism to enhance for high degree of spatial resolution that is a characteristic of high optical harmonics, and so afford a significantly increased precision of localization, to expedite coupling to nanostructures and waveguides. The production of individual 'needle beams' of potentially sub-micron width compares favourably with some of the best, advanced methods of producing such beams,[37] with sub-wavelength dimensions now in prospect.

We thank Professor Robert Lipson, at the University of Victoria, Canada, and also Dr David Bradshaw and Dr Kayn Forbes at the University of East Anglia, Norwich, U.K., for helpful and encouraging comments on this work.